\documentclass[12pt]{article}
\usepackage{times}
\usepackage{geometry}
\usepackage[utf8]{inputenc}
\usepackage{enumitem,amssymb,graphicx,sidecap}
\usepackage{ragged2e}
\newlist{thematic}{itemize}{8}
\setlist[thematic]{label=$\square$}
\usepackage{pifont}

\begin{document}
\raggedright
\huge
Astro2020 Science White Paper \linebreak

Mapping the Inner Structure of Quasars with Time-Domain Spectroscopy \linebreak
\normalsize

\noindent \textbf{Thematic Areas:} \hspace*{60pt} $\square$ Planetary Systems \hspace*{10pt} $\square$ Star and Planet Formation \hspace*{20pt}\linebreak
$\boxtimes$ Formation and Evolution of Compact Objects \hspace*{31pt} $\square$ Cosmology and Fundamental Physics \linebreak
  $\square$  Stars and Stellar Evolution \hspace*{1pt} $\square$ Resolved Stellar Populations and their Environments \hspace*{40pt} \linebreak
  $\boxtimes$    Galaxy Evolution   \hspace*{45pt} $\square$             Multi-Messenger Astronomy and Astrophysics \hspace*{65pt} \linebreak
  
\textbf{Principal Author:}

Name:	Yue Shen
 \linebreak						
Institution:  University of Illinois at Urbana-Champaign
 \linebreak
Email: shenyue@illinois.edu
 \linebreak
Phone:  217-265-4072
 \linebreak
 
\textbf{Co-authors:} (names and institutions)
  \linebreak
 Scott Anderson, University of Washington, USA\\
 Edo Berger, Harvard-Smithsonian Center for Astrophysics, USA\\
 W.N. Brandt, The Pennsylvania State University, USA\\
 Gisella De Rosa, Space Telescope Science Institute, USA\\
 Xiaohui Fan, University of Arizona, USA\\
 Laura Ferrarese, National Research Council of Canada, Canada\\
 Suvi Gezari, University of Maryland, USA\\
 Matthew Graham, California Institute of Technology, USA\\
 Jenny Greene, Princeton University, USA\\
 Catherine J. Grier, University of Arizona, USA \\
 Josh Grindlay, Harvard University, USA\\
 Daryl Haggard, McGill University, Canada\\
 Patrick B. Hall, York University, Canada\\
 Luis Ho, Kavli Institute for Astronomy and Astrophysics, Peking University, China\\
 Hector Ibarra Medel, University of Illinois at Urbana-Champaign, USA\\
 Dragana Ilic, University of Belgrade, Serbia\\
 Zeljko Ivezic, University of Washington, USA\\
 Jacob Jencson, California Institute of Technology, USA\\
 Linhua Jiang, Kavli Institute for Astronomy and Astrophysics, Peking University, China\\
 St\'ephanie Juneau, National Optical Astronomy Observatory, USA\\
 Mansi Kasliwal, California Institute of Technology, USA\\
 Juna Kollmeier, Carnegie Observatories, USA\\
 Alexander Kutyrev, NASA/GSFC, USA\\
 Jennifer I-Hsiu Li, University of Illinois at Urbana-Champaign, USA\\
 Guilin Liu, University of Science and Technology of China, China\\
 Xin Liu, University of Illinois at Urbana-Champaign, USA\\
 Chelsea MacLeod, Harvard-Smithsonian Center for Astrophysics, USA\\
 Gary Melnick, Harvard-Smithsonian Center for Astrophysics, USA\\
 Brian Metzger, Columbia University, USA\\
 Adam D. Myers, University of Wyoming, USA\\
 Christopher O'Dea, University of Manitoba, Canada\\
 Andreea Petric, Institute for Astronomy, University of Hawaii, USA\\
 Luka \v{C}. Popovi\'c, Astronomical Observatory Belgrade, Serbia\\
 Abhishek Prakash, IPAC, California Institute of Technology, USA \\
 Bill Purcell, Ball Aerospace, USA\\
 Gordon T. Richards, Drexel University, USA\\
 George Rieke, University of Arizona, USA\\
 Nial Tanvir, Leicester University, UK\\
 Benny Trakhtenbrot, Tel Aviv University, Israel \\
 Michael Wood-Vasey, University of Pittsburgh, USA\\
 Yongquan Xue, University of Science and Technology of China, China\\
 Qian Yang, University of Illinois at Urbana-Champaign, USA
 \linebreak

\textbf{Abstract  (optional):} The ubiquitous variability of quasars across a wide range of wavelengths and timescales encodes critical information about the structure and dynamics of the circumnuclear emitting regions that are too small to be directly resolved, as well as detailed underlying physics of accretion and feedback processes in these active supermassive black holes. We emphasize the importance of studying quasar variability with time-domain spectroscopy, focusing on two science cases: (1) reverberation mapping (RM) to measure the broad-line region sizes and black hole masses in distant quasars; (2) spectroscopic follow-up of extreme variability quasars that dramatically change their continuum and broad-line flux within several years. We highlight the need for dedicated optical-infrared spectroscopic survey facilities in the coming decades to accompany wide-area time-domain imaging surveys, including: (1) the next phase of the Sloan Digital Sky Survey (SDSS-V; $\sim 2020-2025$), an all-sky, time-domain multi-object spectroscopic survey with 2.5m-class telescopes; (2) the planned Maunakea Spectroscopic Explorer, a dedicated 10m-class spectroscopic survey telescope with a $1.5\,{\rm deg}^2$ field-of-view and multiplex of thousands of fibers in both optical and near-IR ($J+H$) to begin operations in 2029; (3) the Time-domain Spectroscopic Observatory (TSO), a proposed Probe-class $\sim 1.3$m telescope at L2, with imaging and spectroscopy ($R = 200, 1800$) in 4 bands ($0.3 - 5\,\mu$m) and rapid slew capability to 90\% of sky, which will extend the coverage of H$\beta$ to $z=8$.

\pagebreak
\setcounter{page}{1}

\section{Introduction}

Quasars (luminous Active Galactic Nuclei, or AGN) are among the most luminous extragalactic objects in the Universe. Powered by rapid mass accretion onto the supermassive black holes (SMBHs) at the center of distant galaxies, quasars are beacons marking growing massive black holes (BH) and their host galaxies to the farthest reach of modern telescopes (e.g., a recently discovered quasar at $z\sim 7.5$, Ba\~{n}ados et~al.\ 2018). The puzzling tight correlations between the mass of these SMBHs and properties of their host galaxies discovered for local dormant BHs (e.g., Kormendy \& Ho 2013, McConnell \& Ma 2013) suggest a possible co-evolution scenario where the feedback from BH accretion regulates the growth of the host bulge, and potentially impacts the circumgalactic environment. Therefore studying quasar phenomena is not only important for understanding BH accretion, but also crucial for understanding galaxy evolution. 
\quad The inner regions of quasars emit most of the luminosity of these accreting BHs, and produce copious diagnostic features in the observed rest-frame UV-infrared spectrum. The basic structure of quasars consists of an accretion disk surrounding the BH that produces the intense ionizing radiation needed to power the broad emission line region (BLR), the dust torus, and the narrow emission line region (NLR), which extends to galactic scales. The inner structures of quasars (the accretion disk, the BLR, and the dust torus) are all within the sphere of influence of the SMBH, where the most important processes associated with accretion are occurring  (e.g., winds and outflows, photoionization, chemical enrichment, dynamical evolution, etc.). Given the compact sizes of these inner regions ({sub-parsec for a $10^{8}\,M_\odot$ BH}), variability induced by instabilities in the accretion flow, reprocessing of light, and dynamical evolution of the inner regions can be observed over a broad range of timescales from intra-hours to decades, {\bf providing observational constraints on these regions that are too small to be spatially resolved directly. }

\quad Variability is a hallmark characteristic of quasars. Nearly all quasars vary across the entire electromagnetic spectrum, with the variability amplitude depending on timescale, wavelength, and accretion parameters (e.g., BH mass and accretion rate). This ubiquitous variability encodes critical information about the structure and dynamics of the emitting regions, as well as detailed underlying physics of accretion and feedback processes in these active SMBHs. 

\quad This White Paper discusses the needs to understand the accretion processes of quasars and to measure their fundamental physical properties using time-domain spectroscopy. Given the richness of this topic, we will focus on two science cases with time-domain spectroscopy of quasars: (1) reverberation mapping to measure the structure of the BLR (and the accretion disk) and to estimate quasar BH masses throughout cosmic epochs; (2) characterization of extremely variable quasars and their implications for SMBH accretion. We conclude with opportunities from future surveys and facilities in the upcoming landscape of time-domain astrophysics. 

\section{Fundamental Questions on SMBH Growth}

Measuring accurate BH masses is essential to SMBH research. In particular, BH masses for quasars at different cosmic epochs and with different luminosities are critical to understanding:

\begin{enumerate}
\item[$\bullet$] BH accretion processes, and the nature of SMBH feedback that may impact the host galaxy and the circumgalactic environment;
\item[$\bullet$] when and how SMBHs and host galaxies establish their observed local scaling relations between BH mass and host properties;
\item[$\bullet$] the formation and seeding scenarios for the first-generation SMBHs observed at cosmic dawn ($z> 6$). 
\end{enumerate}
To constrain the cosmic assembly of SMBHs we also need to measure a large number of SMBH masses at any given redshift, requiring dedicated survey resources.  

\quad Dynamical BH masses with spatially resolved gas and stellar kinematics are only possible in the nearby universe even with future large-aperture telescopes, and are predominantly applicable to inactive SMBHs (e.g., Kormendy \& Ho 2013), where the measurements are not complicated by contamination from AGN light. Beyond redshift $z\sim 0.3$, the sphere of influence of all but the most massive and elusive $>10^{10}\,M_\odot$ SMBHs becomes unresolvable, and the prime method to measure these distant SMBH masses in the foreseeable future is reverberation mapping (RM, e.g., Blandford \& McKee 1982, Peterson 1993) of unobscured broad-line quasars. RM measures a {characteristic size} of the BLR -- the gaseous region deep within the gravitational potential of the BH producing the broad emission lines -- from the time lag (i.e., light echo) between variability of the continuum emission, which powers the BLR, and the delayed response of the BLR emission. The mass of the SMBH can then be derived by combining this size with the velocity dispersion of the BLR inferred from the width of the broad lines using the virial theorem. More sophisticated modeling of the velocity-resolved RM data can further constrain the detailed structure and kinematics of the BLR (e.g., Pancoast et~al.\ 2012, Grier et~al.\ 2013). Importantly, local RM results anchor the so-called ``single-epoch virial BH mass estimators'' (e.g., Vestergaard \& Peterson 2006) that provide an empirical BH mass estimate using single-epoch spectroscopy, which have been widely applied to quasars at all redshifts and luminosities. However, due to the time-consuming and resource-intensive nature of RM, these local results only include $\sim 60$ AGN (almost exclusively at $z<0.3$) with robust lag measurements, most of which are for the broad H$\beta$ line only (e.g., Bentz et al. 2013) and for low-luminosity AGN as opposed to luminous quasars. Given the small size of the local RM sample and their intrinsic differences from distant quasars, significant (but not well quantified) systematic uncertainties may affect these single-epoch methods, a situation that desperately needs improvement (for a detailed discussion on the limitations of these quasar BH mass recipes, see, e.g., Shen 2013). 

\section{Reverberation Mapping Beyond 2020}

The only way to improve BH weighing methods for distant quasars is by expanding substantially the sample of objects with direct RM measurements, to improve the statistics, and more importantly, to cover the parameter space occupied by the general quasar population. However, traditional, single-object RM is inefficient in expanding the sample size and covering the parameter space of AGN. Spectroscopic RM with Multi-Object Spectrographs (hereafter {\bf MOS-RM}) has recently emerged as a new approach to perform efficient RM for large samples of distant quasars (e.g., Shen et~al.\ 2015, King et~al. 2015, Grier et~al.\ 2017), taking advantage of the multiplexing power and wide field-of-view of these MOS facilities. MOS-RM is an important step to extend the RM exercise to the distant universe and for uniform quasar samples. With orders of magnitude increase in the RM sample statistics, and expansion of the sample in luminosity and redshift to cover the rise and fall of the cosmic quasar population between $1\lesssim z\lesssim 6$, we will be able to: (1) derive reliable RM-based BH mass measurements for quasars up to $z\sim 6$ and enable better calibrations of single-epoch BH mass estimators for the general quasar population; (2) provide the basis for studying the evolution of the scaling relations between SMBH mass and host properties; and (3) greatly improve our understanding of the structure and dynamics of the BLR and accretion disk from echo mapping of different broad line species and the underlying continuum. 

\quad The basic operational concept and expected yields of a MOS-RM program are summarized in {\bf Figure 1}. In a MOS-RM program, all quasars within the field-of-view are spectroscopically monitored simultaneously. The cadence is designed to vary between dense sampling (for short lags) and sparse sampling (for long lags) throughout a typical baseline of several years in order to detect the longest lags possible. The expected yields are functions of the baseline, SNR, and cadence of the program, which can be estimated with detailed simulations (e.g., Shen et~al. 2015). Such MOS-RM programs are usually accompanied by dense photometric monitoring either from public surveys (e.g., LSST) or dedicated imaging programs, where the photometric light curves provide better sampling of the driving continuum variability.  

\quad We also include the forecast for TSO in {\bf Figure 1}. TSO is not a MOS facility, but its rapid slew capability over most of the sky enables efficient, targeted monitoring of quasars with variability triggers from, e.g., LSST. This observing strategy allows TSO to be used as a powerful facility for quasar RM. More importantly, TSO will provide much broader spectral coverage ($0.3-5\,\mu$m) to allow H$\beta$ RM at $z=8$ (with sufficient time baseline from early LSST triggering). 

\begin{figure*}[!h]
\centering
\includegraphics[width=0.8\textwidth]{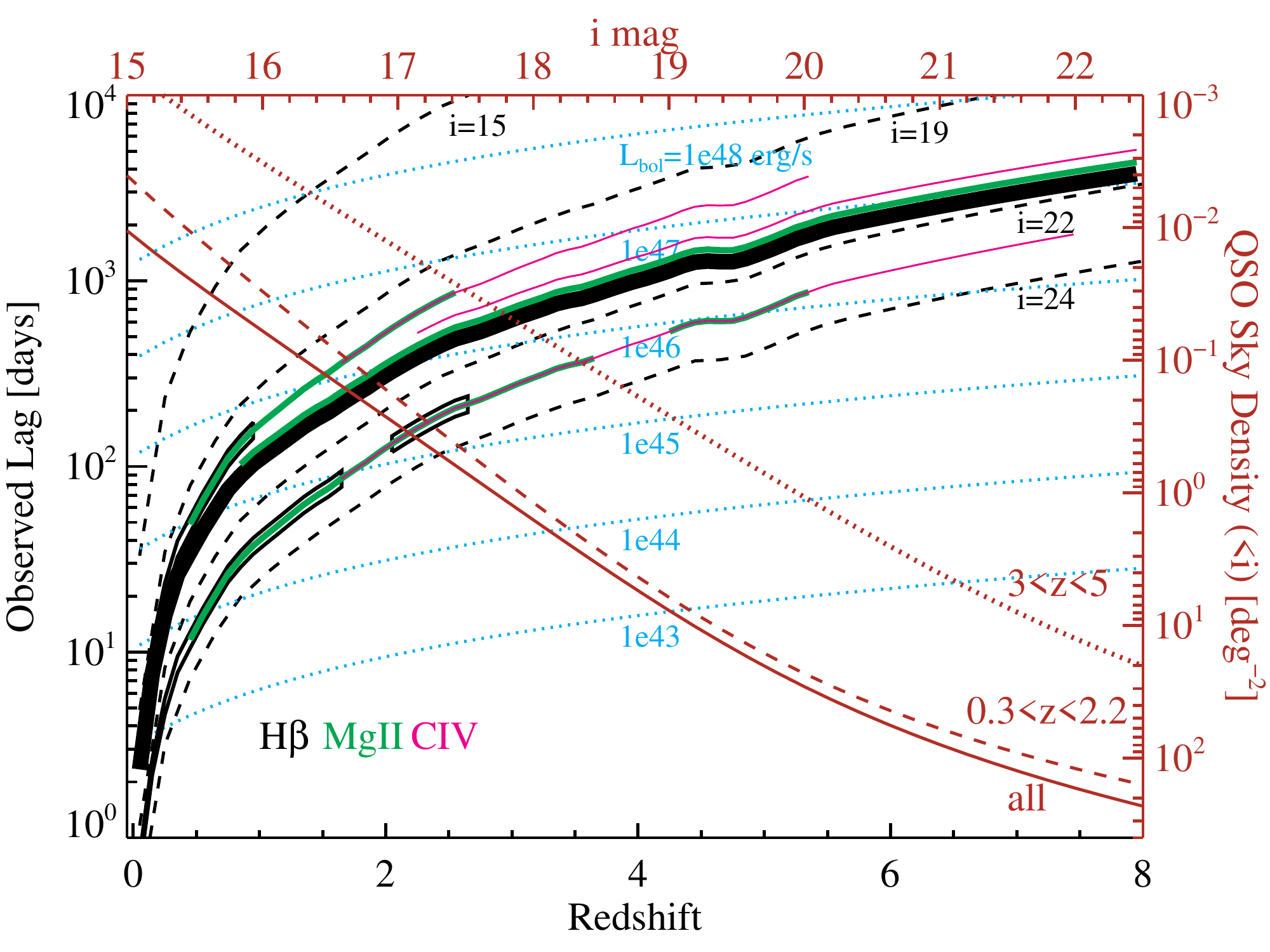}
\caption{\footnotesize Lag detection capabilities with dedicated upcoming RM programs targeting quasars beyond the local universe. The left vertical axis is the expected lag in the observed frame against redshift (bottom axis). The blue dotted lines show the approximate correspondences to constant quasar bolometric luminosities, converted using the $R-L$ relation in Bentz et~al. (2013) with a $40\%$ scatter. The black dashed lines show the apparent $i$-band magnitudes for $i=15, 19, 22, 24$. The colored line segments denote the {\em shortest} measurable lags for H$\beta$ (black), MgII (green) and CIV (magenta), the three major broad lines for RM. The upper three-color branch corresponds to the limit for the Black Hole Mapper component in the upcoming SDSS-V survey (Kollmeier et al. 2017). The middle three-color branch corresponds to a proposed concept of the space-based TSO mission matched to the same SNR (${\rm SN}\approx 10$ per resolution element) and spectral resolution ($R\sim 2000$). The lower three-color branch corresponds to the MOS-RM program proposed for the Maunakea Spectroscopic Explorer (MSE, McConnachie et~al. 2016). The difference in broad line coverage reflects the different spectral coverage from these planned projects. {\bf The top/right axes and the brown lines indicate the sky density of quasars} at different limiting $i$-band magnitude, e.g., there are $\sim 200$ broad-line quasars per ${\rm deg^{2}}$ at $i<22$. }
\label{fig:mosrm}
\end{figure*}

\quad In addition to the main RM science, a wealth of parallel science cases will be enabled by the time-resolved and multi-wavelength follow-up data from these programs. These parallel science cases range from rare, extreme variability of individual quasars over multi-year timescales (see \S4), to detailed host galaxy studies using the extremely deep coadded spectroscopy from the multi-epoch RM data. For example, a typical RM monitoring program will re-visit the same target for $\sim 100$ times or more, and the coadded depth is often sufficient to detect host galaxy features within the quasar spectrum (e.g., Shen et al. 2015b, Matsuoka et al. 2015). This provides an ideal case to study the correlation between quasar BH mass (from RM) and host galaxy properties at high redshifts. Another example is UV-optical continuum reverberation mapping (e.g., Fausnaugh et~al.\ 2016), which constrains the size of the accretion disk by measuring intra-band time lags from dense photometric or spectroscopic light curves from the same RM program. 

\section{Changing-look Quasars}

Time-domain spectroscopy of quasars in general also promises to deliver additional important clues to accretion processes. A rare phenomenon of AGN in which the continuum and broad line flux varies by more than a factor of a few within a few years, first discovered decades ago (e.g., Penston \& Perez 1984, Goodrich 1995), has regained interest from recent time-domain imaging and spectroscopic programs (e.g., LaMassa et~al. 2015, Runnoe et~al. 2016, MacLeod et~al. 2018, Yang et~al. 2018). In extreme cases, the quasar completely changes spectroscopic types, from a broad-line Type 1 object to a narrow-line Type 2 object and vice versa (see {\bf Figure 2} for an example). Follow-up studies of these systems concluded that a dramatic change in the accretion flow, rather than changes in the obscuration configuration, is the most likely explanation of the changes in their spectral appearance (Runnoe et~al. 2016, Sheng et~al. 2017). This population of Changing-Look (CL) quasars not only challenges the canonical unified model of AGN, but also is difficult to understand given that the observed timescale of changes (less than a few years in the quasar rest frame) is much shorter than the viscous timescale required to change the accretion rate significantly in the standard accretion disk theory (see discussions in, e.g., Lawrence 2018).

\quad At the extreme end of quasar variability, CL quasars provide an exciting new probe of the structure of quasar accretion disks and their nuclear environments, allowing probes of (1) the radial structure of the accretion disk, by following the change in the continuum emission in spectroscopic monitoring during a CL quasar transition; (2) the structure and excitation mechanisms of the BLR, through forward modeling of the response of the broad emission lines to the variability in the continuum emission; (3) the relation between the black hole mass and galaxy stellar velocity dispersion and stellar mass (the $M_\mathrm{BH}$ -- $\sigma_*$ and $M_\mathrm{BH}$--$M_*$ relations), through measurements of the $M_\mathrm{BH}$ using broad lines during the bright active state (quasar light dominant) and $\sigma_*$ ($M_*$) during the dim inactive state (galaxy light dominant).

\quad CL quasars are now being discovered systematically from repeat spectroscopic surveys such as the Time Domain Spectroscopic Survey in SDSS-IV, and by photometric monitoring and follow-up optical spectroscopy of quasars (galaxies) whose nuclei dramatically dim (brighten). The sample of CL quasars is growing (currently tens of cases robustly confirmed), but high-cadence spectroscopic observations of the transition are still lacking due to limited observing resources and the rareness of the phenomenon. LSST will systematically unearth thousands of CL quasars {\em as they transition}, allowing detailed tracking of the dynamical response of the accretion disk and broad line region in dedicated spectroscopic follow-up programs.  


\begin{SCfigure}
\centering
\includegraphics[width=0.6\textwidth]{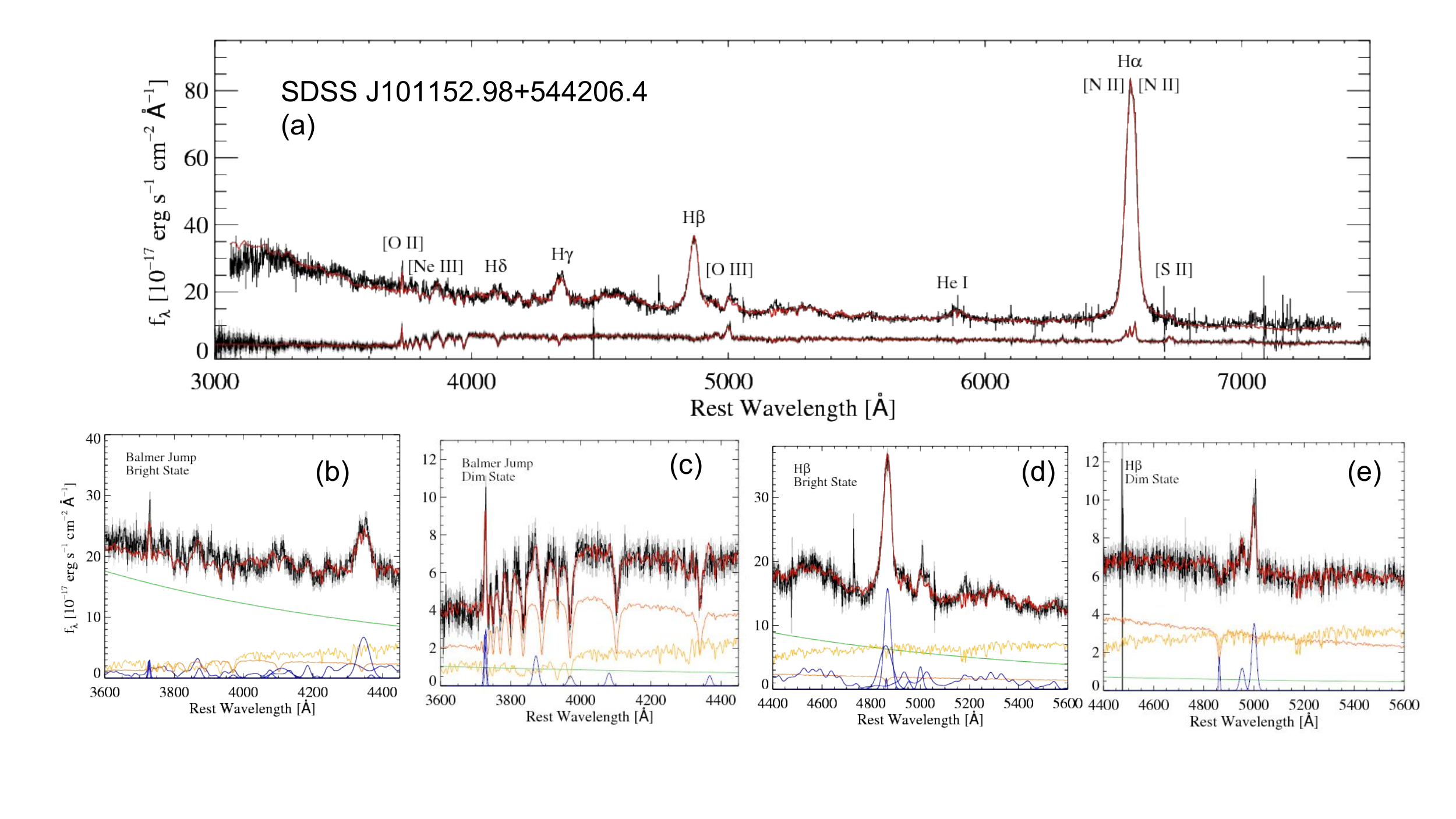}
\caption{\footnotesize Spectral comparison of two epochs for the ``changing-look'' quasar, SDSS J101152.98+544206.4 (Runnoe et al., 2016). This $z=0.246$ quasar dimmed by $\sim 2.5$ mag over about 9 yrs in the rest frame; between the bright (01-13-2003) and the dim state (02-20-2015), a strong decrease in the blue continuum is seen, accompanied by a similar decrease in the broad line flux. Spectral decomposition is shown for the Balmer jump region for bright (b) and dim (c) epochs, and for the H$\beta$-[OIII] region for bright (d) and dim (e) epochs. }
\label{fig:clq}
\end{SCfigure}

\section{New Surveys and New Facilities}

Both science cases (reverberation mapping and changing-look quasars) will benefit from dedicated, spectroscopic monitoring programs from future MOS surveys. SDSS-V (2020-2025) is the next generation of the Sloan Digital Sky Survey (Kollmeier et al. 2017), and includes a Black Hole Mapper component which will perform MOS-RM for over a thousand quasars and obtain repeat optical spectroscopy for $>20,000$ quasars to systematically discover CL quasars. SDSS-V will significantly advance time-domain spectroscopy of quasars by 2025, and pave the way for later, more ambitious programs. The Maunakea Spectroscopic Explorer (MSE) is a planned 10-m class telescope$+$MOS with a $1.5\,{\rm deg}^2$ field-of-view and multiplex of thousands of fibers in both optical and near-IR (McConnachie et al. 2016), planned to start operation in 2029. MSE will significantly improve the depth and redshift range of H$\beta$ coverage (see {\bf Figure 1}) for MOS-RM programs beyond $\sim 2030$. In addition to wide-field MOS facilities, new survey facilities and instruments with broader wavelength coverage into the mid-infrared will be extremely useful to observe rest-frame optical lines (such as the Balmer lines) at high redshift. For example, a Probe-class NASA mission concept, the Time-domain Spectroscopic Observatory (TSO; PI J. Grindlay), is being developed to provide imaging and spectroscopy in 4 bands,  from $0.3$ to $5\,\mu $m, and spectral resolution $R=200$ and 1800. By targeting variable quasars selected from LSST, TSO can provide efficient RM monitoring in H$\beta$ out to $z=8$, and high-cadence spectroscopic follow-up of Changing-look transitions of quasars, with a much more complete coverage of restframe UV-optical lines than ground-based optical/IR spectroscopy.

\pagebreak
\setcounter{page}{1}
\textbf{References}

Ba\~{n}ados, E., et~al.\ 2018, Nature, 553, 473\\
Bentz, M.~C., et al.\ 2013, ApJ, 767, 149\\
Blandford, R.~D., \& McKee, C.~F.\ 1982, ApJ, 255, 419\\
Fausnaugh, M.~M., et~al.\ 2016, ApJ, 821, 56\\
Goodrich, R.~W.\ 1995, ApJ, 440, 141\\
Grier, C.~J., et~al.\ 2013, ApJ, 764, 47\\
Grier, C.~J., et~al.\ 2017, ApJ, 851, 21\\
Kollmeier, J.~A., et~al.\ 2017, arXiv:1711.03234\\
Kormendy, J. \& Ho, L.~C.\ 2013, ARA\&A, 51, 511\\
King, A.~L., et~al.\ 2015, MNRAS, 453, 1701\\
LaMassa, S.~M., et~al.\ 2015, ApJ, 800, 144\\
Lawrence, A.\ 2018, Nature Astronomy, 2, 102\\
MacLeod, C.~L., et~al.\ 2018, ApJ, in press, arXiv:1810.00087\\
Matsuoka, Y., et~al.\ 2015, ApJ, 811, 91\\ 
McConnachie, A., et~al.\ 2016, arXiv:1606.00043\\
McConnell, N.~J., \& Ma., C.-P.\ 2013, ApJ, 764, 184\\ 
Pancoast, A., et~al.\ 2012, ApJ, 754, 49\\
Penston, M. V., \& Perez, E.\ 1984, MNRAS, 211, 33\\
Peterson, B.~M., et~al.\ 1993, PASP, 105, 247\\
Runnoe, J.~C., et~al.\ 2016, MNRAS, 455, 1691\\
Shen, Y.\ 2013, BASI, 41, 61\\
Shen, Y., et~al.\ 2015, ApJS, 216, 4\\
Shen, Y., et~al.\ 2015b, ApJ, 805, 96\\
Sheng, Z., et~al.\ 2017, ApJL, 846, L7\\
Vestergaard, M., Peterson, B.~M.\ 2006, ApJ, 641, 689\\
Yang, Q., et~al. 2018, ApJ, 862, 109


\end{document}